\documentclass[fleqn,usenatbib]{mnras}
\usepackage{amsmath,amssymb}
\usepackage{mathtools}
\usepackage{graphicx}
\DeclareGraphicsExtensions{.png,.pdf}
\usepackage{txfonts}
\usepackage[english]{babel}
\usepackage[utf8]{inputenc}
\usepackage[T1]{fontenc}
\usepackage{cuted}
\usepackage{soul}
\allowdisplaybreaks
\usepackage{comment}
\usepackage{multirow}
\usepackage{xcolor}

\newcommand\rhocen{\rho_{150}}
\newcommand\rperi{r_{\rm p}}
\newcommand\Msun{{\rm M}_{\odot}}

\newcommand{\real}[1]{{#1}^{\ast}}

\newcommand{\orcid}[1]{\href{https://orcid.org/#1}{\,\includegraphics[height=\fontcharht\font`\B]{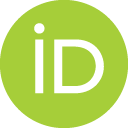}}}

\newcommand{\beq}{\begin{equation}}
\newcommand{\eeq}{\end{equation}} 

\renewcommand\bphi{b_{\phi}}

\newcommand\B{\mathcal{B}}

\defcitealias{Hay20}{H20}
\newcommand{\HAY}{\citetalias{Hay20}}

\defcitealias{Kap19}{K19}
\newcommand{\KAP}{\citetalias{Kap19}}

\defcitealias{Rea19}{R19}
\newcommand{\READ}{\citetalias{Rea19}}

\defcitealias{Pac22}{P22}
\newcommand{\PAC}{\citetalias{Pac22}}

\defcitealias{Bat22}{B22}
\newcommand{\BAT}{\citetalias{Bat22}}

\defcitealias{Fri18}{F18}
\newcommand{\FRI}{\citetalias{Fri18}}

\begin{document}

\date{Draft, April 6, 2023}

\title[Are $\rperi$ and $\rhocen$ of MW dSphs anti-correlated?]{On the anti-correlation between pericentric distance and inner dark matter density of Milky Way's  dwarf spheroidal galaxies}

\author[S.\ Cardona-Barrero et al.]{\parbox{\textwidth}{Salvador Cardona-Barrero$^{1,2}$\thanks{E-mail:scardona@iac.es}$^{\rm \orcid{0000-0002-9990-4055}}$, 
Giuseppina Battaglia$^{1,2}$$^{\rm \orcid{0000-0002-6551-4294}}$, 
Carlo Nipoti$^{3}$$^{\rm \orcid{0000-0003-3121-6616}}$,\\ 
Arianna Di Cintio$^{2,1}$$^{\rm \orcid{0000-0002-9856-1943}}$ \\} \\
  \parbox{\textwidth}{
$^{1}$ Instituto de Astrof\'isica de Canarias, Calle V\'ia L\'actea s/n, E-38206 La Laguna, Tenerife, Spain \\
$^{2}$ Universidad de La Laguna Avda. Astrof\'isico Fco. S\'anchez, E-38205 La Laguna, Tenerife, Spain \\
$^3$Dipartimento di Fisica e Astronomia ``Augusto Righi'', Universit\`a di Bologna,
    via Gobetti 93/2, I-40129 Bologna, Italy \\
}}

\maketitle 

\begin{abstract}
\noindent An anti-correlation between the central  density of the dark matter halo ($\rho_{150,\ {\rm DM}}$) and the pericentric distances ($\rperi$) of the Milky Way's (MW's) dwarf spheroidal galaxies (dSphs)  has been reported in the literature. The existence and origin of such anti-correlation is however controversial, one possibility being that only the densest dSphs can survive the tidal field towards the centre of our  Galaxy.
In this work, we place particular emphasis on quantifying the statistical significance of such anti-correlation, by using  available  literature data in order to explore its robustness under different assumptions on the MW gravitational potential, and for various derivations of  $\rhocen$  and $\rperi$. We consider models in which the MW is isolated and has a low  ($8.8\times10^{11}\,\Msun$) and  high ($1.6\times10^{12}\, \Msun$) halo mass, respectively, as well as configurations in which the MW's potential is perturbed by a Large Magellanic Cloud (LMC) infall.
We find that, while data  generally support models in which the dSphs' central DM density decreases as a function of their pericentric radius, this anti-correlation is statistically significant at $3\sigma$ level only in $\sim$12$\%$ of the combinations of $\rhocen$ and $\rperi$ explored. Moreover, including the impact of the LMC's infall onto the MW weakens or even washes away this anti-correlation, with respect to  models in which the MW is isolated.
Our results suggest that the strength and existence of such
anti-correlation is still debatable: exploring it with high-resolution simulations including baryonic physics and different DM flavours will help us to understand its emergence.
\end{abstract}

\begin{keywords}
galaxies: Local Group --
galaxies: dwarf --
galaxies: kinematics and dynamics --
galaxies: dark matter
\end{keywords}

\section{Introduction}
\label{sec:intro}

Thanks to several observational campaigns and theoretical work on
dynamical modelling, the dark matter (DM) content and orbital parameters of the Milky Way's (MW) dwarf spheroidal galaxies (dSphs)\footnote{Following \cite{Sim19}, we adopt the nomenclature 'dwarf spheroidal galaxy'/'ultra faint dwarf' for the galaxies brighter/fainter than absolute V-band magnitude M$_V= -7.7$.} are now relatively well known \citep[e.g.][and references therein]{BN22,Bat22}.  The inner DM density of dSphs is often quantified by measuring $\rhocen$, defined as the DM density at a distance of $150\,$pc from the
centre of the dwarf \citep[e.g.][hereafter \READ]{Rea19}.  The orbit of each dSph,
which depends on the gravitational potential assumed for the MW, can be described by different parameters, among which the
pericentric radius $\rperi$ (minimum distance of the dwarf centre of
mass from the Galactic centre), which gives an indication of the importance of tidal effects.

\begin{table*}
    \centering
    \begin{tabular}{l|cccccccc} 
    \hline
 & Draco & UMi & Carina & Sextans & LeoI & LeoII & Sculptor & Fornax \\ \hline
 \multicolumn{9}{c}{Pericentres: $\rperi~\left[{\rm kpc}\right]$} \\ \hline
F18 & ${42.0}_{+11.0}^{+16.0}$ & ${44.0}_{+10.0}^{+12.0}$ & ${103.0}_{+23.0}^{+8.0}$ & ${79.0}_{+8.0}^{+9.0}$ & ${63.0}_{+47.0}^{+221.0}$ & ${67.0}_{+52.0}^{+154.0}$ & ${69.0}_{+9.0}^{+10.0}$ & ${100.0}_{+33.0}^{+28.0}$ \\  
B22\_L & ${51.7}_{+6.0}^{+4.0}$ & ${48.9}_{+3.0}^{+3.0}$ & ${106.7}_{+5.0}^{+6.0}$ & ${74.5}_{+6.0}^{+4.0}$ & ${46.6}_{+26.0}^{+30.0}$ & ${115.5}_{+59.0}^{+88.0}$ & ${63.6}_{+3.0}^{+4.0}$ & ${89.4}_{+26.0}^{+31.0}$ \\  
B22\_H & ${37.6}_{+4.0}^{+4.0}$ & ${34.9}_{+3.0}^{+3.0}$ & ${102.8}_{+32.0}^{+10.0}$ & ${64.0}_{+6.5}^{+5.0}$ & ${35.0}_{+20.0}^{+24.0}$ & ${69.0}_{+29.0}^{+64.0}$ & ${48.7}_{+4.0}^{+4.0}$ & ${56.2}_{+15.0}^{+22.0}$ \\  
B22\_LMC & ${100.0}_{+19.0}^{+22.0}$ & ${70.4}_{+5.0}^{+8.5}$ & ${98.0}_{+24.0}^{+14.0}$ & ${71.6}_{+6.0}^{+4.5}$ & ${40.1}_{+24.0}^{+29.0}$ & ${105.4}_{+50.0}^{+115.0}$ & ${47.7}_{+3.4}^{+3.4}$ & ${91.7}_{+25.0}^{+32.0}$ \\  
P22 & ${40.4}_{+5.4}^{+6.5}$ & ${41.8}_{+4.5}^{+5.3}$ & ${114.4}_{+11.8}^{+49.7}$ & ${82.8}_{+4.0}^{+3.7}$ & ${42.9}_{+23.2}^{+28.9}$ & ${54.3}_{+31.6}^{+55.7}$ & ${55.0}_{+5.2}^{+5.5}$ & ${85.2}_{+29.3}^{+38.6}$ \\  
P22\_LMC & ${58.0}_{+9.5}^{+11.4}$ & ${55.7}_{+7.0}^{+8.4}$ & ${77.9}_{+17.9}^{+24.1}$ & ${82.2}_{+4.3}^{+3.8}$ & ${47.5}_{+24.0}^{+30.9}$ & ${61.4}_{+34.7}^{+62.3}$ & ${44.9}_{+3.9}^{+4.3}$ & ${76.7}_{+27.0}^{+43.1}$ \\  
\hline  \multicolumn{9}{c}{Central Densities:  $\rhocen ~\left[10^7~\Msun~{\rm kpc}^{-3}\right]$} \\ \hline
R19 & ${23.6}_{+2.9}^{+2.9}$ & ${15.3}_{+3.2}^{+3.5}$ & ${11.6}_{+2.2}^{+2.0}$ & ${12.8}_{+2.9}^{+3.4}$ & ${17.7}_{+3.4}^{+3.3}$ & ${18.4}_{+1.6}^{+1.7}$ & ${14.9}_{+2.3}^{+2.8}$ & ${7.9}_{+1.9}^{+2.7}$ \\  
H20 & ${23.5}_{+6.3}^{+12.8}$ & ${23.8}_{+7.2}^{+38.6}$ & ${10.9}_{+3.2}^{+8.2}$ & ${5.2}_{+2.3}^{+3.6}$ & ${26.4}_{+9.1}^{+22.3}$ & ${20.2}_{+6.1}^{+12.7}$ & ${21.4}_{+6.3}^{+12.6}$ & ${12.2}_{+2.3}^{+3.2}$ \\  
K19\_NFW & ${21.7}_{+2.2}^{+2.7}$ & ${25.2}_{+4.5}^{+2.9}$ & ${10.3}_{+0.9}^{+1.1}$ & ${11.0}_{+1.8}^{+2.9}$ & ${15.1}_{+2.4}^{+3.4}$ & ${17.1}_{+3.8}^{+2.4}$ & ${17.1}_{+2.2}^{+2.1}$ & ${7.5}_{+1.4}^{+2.0}$ \\  
K19\_ISO & ${21.3}_{+4.7}^{+5.4}$ & ${25.4}_{+5.7}^{+6.1}$ & ${5.7}_{+1.7}^{+3.2}$ & ${8.5}_{+3.5}^{+5.0}$ & ${14.1}_{+4.5}^{+5.5}$ & ${13.5}_{+1.7}^{+4.2}$ & ${16.1}_{+3.3}^{+2.9}$ & ${3.4}_{+1.3}^{+1.7}$ \\ 
\hline
    \end{tabular}
    \caption{Values of the pericentric radius and central DM density of each dSph (columns) compiled from the literature (rows); we refer the reader to Sect.~\ref{sec:dataset} for the labeling of each model. $\rhocen$ and $\rperi$ are given as the $50^\mathrm{th}$ percentiles of the distribution of values, and the lower and upper error-bars bracket the $16^\mathrm{th}$ and $84^\mathrm{th}$ percentiles.}
    \label{tab:sample}
\end{table*}

Using pericentres inferred from the second data release (DR2) from the {\em Gaia} mission \citep{prusti2016gaia, GaiaDR22018}, \citet[][hereafter \KAP]{Kap19} claimed that a sample of nine MW dSphs exhibits an anti-correlation between $\rhocen$ and $\rperi$.  This anti-correlation, which is also found for dark-matter sub-haloes in some cosmological simulations \citep{Rob21,Gen22}, might be a consequence of survivor bias, i.e.\  the fact that lower-density satellites on small pericentre orbits have not survived the tidal field of the MW \citep{Hay20,Gen22}. 
Alternatively, it may be a signature of Self-Interacting Dark Matter (SIDM), as gravothermal core collapse \citep{Balberg2002} is accelerated in dwarfs that undergo tidal stripping \citep{Nishikawa2020}, leading to larger central densities. 

The existence and the strength of such anti-correlation for MW satellite galaxies is a matter of debate \citep{Hay20,Gen22,Hay22}. There is no evidence of anti-correlation when samples of ultra faint dwarfs (UFDs) are analyzed (\KAP{}).  It is also to be considered that while the mass (and average density) of these pressure-supported galaxies is determined with the highest precision within the half-light radius \citep[e.g.][]{Wol10} or $1.8$ times the half-light radius \citep{Err18}, where the mass-anisotropy degeneracy is minimized, mass and density estimates at other locations carry larger uncertainties.

In this paper we revisit the question of the possible anti-correlation between $\rhocen$ and $\rperi$ of MW dSphs with a quantitative approach, by performing a systematic statistical analysis. In particular, we address the question of whether the result is sensitive to the set of literature estimates of $\rhocen$ and $\rperi$ considered. For the latter quantity,  several new determinations have recently been obtained using the more accurate and precise data from the {\it Gaia} early third data release (eDR3) \citep{GaiaeDR32021} and also taking into account self-consistently the impact of the infall of a massive Large Magellanic Cloud (LMC) onto the MW, which can strongly affect the orbital history of MW dSphs \citep[e.g.][]{Patel2020, Bat22, Pac22}. This article is structured in the following way: in Sect.~\ref{sec:dataset} we introduce the set of MW dSphs considered, and the sets of literature estimates for $\rhocen$ and $\rperi$; in Sect.~\ref{sec:Analysis} we present the statistical approaches undertaken, including a method of general validity for data sets with asymmetric error-bars, and test them on mock data sets; in Sect.~\ref{sec:results} we discuss our results and present our conclusions in Sect.~\ref{sec:conclusions}.

\section{Sample and data sets}
\label{sec:dataset}

\begin{figure*}
    \centering
    \includegraphics[width=1\textwidth]{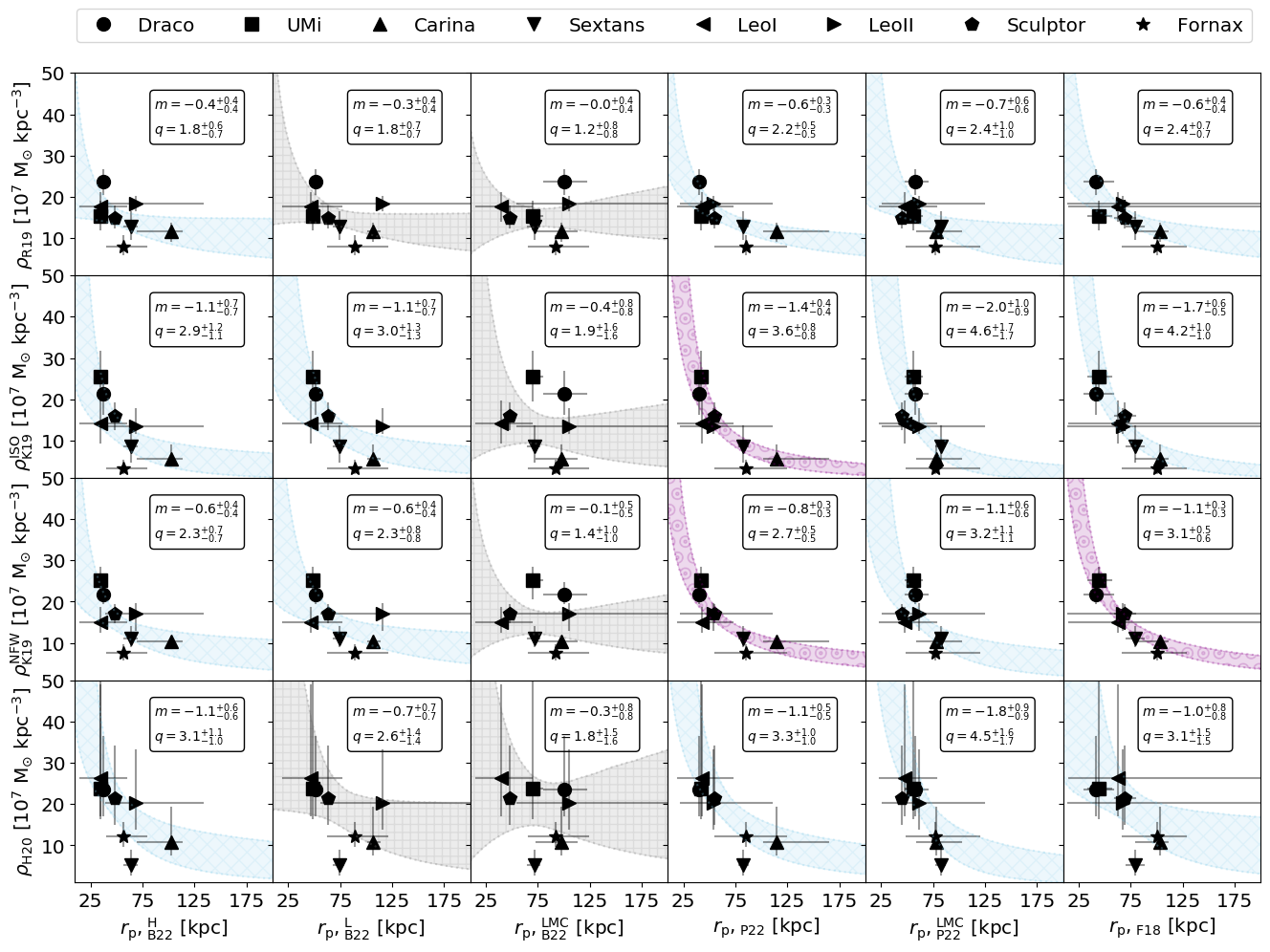}
    \caption{Central dark matter density versus pericentric distance for $8$ classical dSphs. From top to bottom we show the the densities as derived in \READ{}, \KAP{}(assuming an isothermal and NFW DM density profile, respectively) and \HAY{}. The pericentres are from (from left to right): 
    \BAT{} assuming a heavy, and light MW and a light MW with the inclusion of the LMC; \PAC{} without and with LMC, and \FRI{} with a light MW. The details of each data set can be found in Sec.~\ref{sec:dataset}. 
    The shaded region depicts the $68$ per cent confidence-level region of the power-law fit between the central DM density and the pericentric distance, obtained with the percentile fitting method (see Appendix \ref{Apx:ThreeOrderStat}). In each panel we show the recovered median logarithmic slope ($m$) and logarithmic zero point ($q$). The uncertainties of the parameters are indicated with the $16$-th and $84$-th percentiles.  The colours of the reported relations have been chosen to represent the statistical significance of the logarithmic slope: whether is compatible with $0$ within $<1\sigma$ (grey), between $1\sigma$ and $3\sigma$ (light-blue) or within $>3\sigma$ (magenta) or more.}
    \label{fig:correlations}
\end{figure*}

We focus on eight of the MW dSphs, specifically the objects in common between the studies of \KAP{}, \READ{} and \citet[][hereafter \HAY{}]{Hay20}. \KAP{} included also Canes Venatici~I but, as we will see later, the exclusion of this system does not change the conclusions on the existence of the anti-correlation between $\rhocen$ and $\rperi$. 

The values of $\rhocen$ are taken from  \KAP, \READ{} and \HAY{}\footnote{While for \READ{}and \HAY{} the values are tabulated, in \KAP{} they are not and we have digitized their Fig.~$2$.}. 
In \READ{} they were determined with GravSPHERE, which solves the spherical Jeans equation for the projected line-of-sight (l.o.s.) velocity dispersion profile of the stellar component and fits also two higher order 'virial shape parameters' \citep{Mer90, Ric14, Rea17}; it uses a non-parametric form for the enclosed mass as a function of radius, M($<r$). Also \KAP{} solved the spherical Jeans equation for the projected l.o.s.\ velocity dispersion profile but considered one 'virial shape parameter'; for the DM halo density profile, they considered separately a NFW \citep{NFW1996} model and a cored isothermal model; therefore we have two sets of $\rhocen$ for the \KAP{} study (which we label K19\_{NFW} and K19\_{ISO}, respectively). In \HAY{}, the mass modeling was performed by solving the axisymmetric Jeans equations for the second moment of the l.o.s. velocity distribution at a given projected $2$D position; the DM halo was modeled with a generalized Hernquist \citep{Hernquist1990} profile, therefore probing cuspy and cored models, and considering non-spherical DM haloes; $\rhocen$ is calculated along the major axis of the dark matter halo, which is assumed to have the same orientation as the stellar component.

We also consider several determinations of the pericentric radii for the MW dSphs: the Gaia DR2-based determinations for a MW of mass $0.8\times10^{12}\,\Msun$ by \citet[][hereafter \FRI]{Fri18}, as used in \KAP{}\footnote{\KAP{} also demonstrate that their results are essentially unchanged when using the values for a twice as massive MW.}; the Gaia eDR3-based pericentric distances by \citet[][hereafter \BAT{}]{Bat22}  in 3 gravitational potentials, two of them with isolated MW of mass $8.8\times10^{11}\,\Msun$ ('Light', hereafter B22\_L) and $1.6\times10^{12}\,\Msun$ ('Heavy', hereafter B22\_H), respectively, and one with a $8.8\times10^{11}\,\Msun$ MW perturbed by a $1.5\times10^{11}\,\Msun$ LMC (hereafter B22\_{LMC}), and the Gaia eDR3 based values by \citet[][hereafter \PAC{}]{Pac22} in an isolated $1.3\times10^{12}\,\Msun$ MW (hereafter, labelled as P22) and in a MW+LMC potential (P22\_{LMC}), having the LMC a mass of $1.38\times10^{11}\,\Msun$.

Table~\ref{tab:sample} and Fig.~\ref{fig:correlations} (black points with error-bars) present all the sets of $\rhocen$ and $\rperi$ used in this work.
As can be gathered from Fig.\ref{fig:correlations} in a qualitative way, the relationship between $\rperi$ and $\rhocen$ varies,  depending on the pericentric radii adopted as well as on the determinations of the central DM densities. In the next sections, we quantify this visual impression with a quantitative statistical analysis.

\section{Statistical analysis}
\label{sec:Analysis}
In this Section we outline the methodology we use to quantify a possible anti-correlation between $\rhocen$ and $\rperi$. 

As in \KAP{}, we model the relationship between $\rhocen$ and $\rperi$ as a power-law (PL) of the form
\begin{equation} 
    \label{eq:PL}
    \log_{10}\left(\frac{\rhocen}{10^7\,{\rm M_{\odot}\,kpc^{-3}}}\right) = q + m\log_{10}\left(\frac{\rperi}{{\rm kpc}}\right), 
\end{equation}
with $q$ and $m$ being the logarithmic zero-point and slope. 

One of the aspects that we wish to take into account is that usually the error bars in the sets of $\rhocen$ and $\rperi$ are highly asymmetric. However this asymmetry has not been considered when fitting Eq.~\eqref{eq:PL} in the literature. For example, when quantifying the relation between $\rhocen$ and $\rperi$, \KAP{} symmetrize the errors by averaging the upper and lower errors, and shift the median to the mid point (see their appendix~C).   
In this work we explore two different methods that retain the information of the asymmetries of the errors.

The  gathered data (Tab.~\ref{tab:sample} and Fig.~\ref{fig:correlations}) are provided in the literature as the median and the $16^\mathrm{th}$ and $84^\mathrm{th}$ percentiles  ($x_{50\mathrm{th}},~x_{16\mathrm{th}},~x_{84\mathrm{th}}$ respectively) of an underlying distribution. 
Our first approach ('simulating errors' method), described in Appendix~\ref{Apx:Resimulation},  consists in re-simulating the error distribution via the reported percentiles: assuming a probability distribution we fit its corresponding cumulative distribution function (CDF) to the percentiles of each pair of pericentre and central density values. Then via random sampling from the fitted probability distribution we can obtain different random realizations of the original data.  
The second approach ('percentile fitting' method), described in Appendix~\ref{Apx:ThreeOrderStat}, consists in treating each percentile as a random variable and model its probability distribution, thus a fully Bayesian approach is possible via order statistics.  
Finally, for comparison, we have also applied the fitting method used in \KAP{}, which we will refer to as 'symmetrized errors' method.

In order to test the performance of these three methods, we applied them to mock data sets, with a radial coverage and noise level mimicking those of the data in the literature. We generate a set of $8$ pericentric distances $\real{\rperi}$ with similar CDF as the one obtained from the pericentres of \PAC{}; the corresponding DM central densities $\real{\rhocen}$ are obtained from Eq.~\ref{eq:PL} assuming $q=3$ and $m=-1$. We introduce noise by simulating $N=20$ samples\footnote{Note that the choice of N=$20$ does not meet the conditions described in Appendix \ref{Apx:ThreeOrderStat}, i.e. $20$ is not a large number of samples and the fraction $P_{i}(N-1)/100$ is not an integer. Thus we may not expect an excellent performance when using the percentiles method.} extracted from a log-normal distribution centreed on $\real{\rperi}$ and $\real{\rhocen}$ with logarithmic variance $\sigma_{\rm LN}^2$. For each data-point we derive the $16^\mathrm{th}$, $50^\mathrm{th}$ and $84^\mathrm{th}$ percentile. 
In this way we are able to obtain samples of mock data with asymmetric errors similar to those in the literature.

\begin{table}
    \centering
    \begin{tabular}{c|c}
    \hline
    Data Set & $\left<\sigma_{\rm data}\right>$  \\ \hline
$\rho_{\rm R19}$ & $0.190 \pm 0.059$  \\ \\[-1em]
$\rho_{\rm H20}$ & $0.458 \pm 0.129$  \\ \\[-1em]
$\rho_{\rm K19}^{\rm NFW}$ & $0.164 \pm 0.045$   \\ \\[-1em]
$\rho_{\rm K19}^{\rm ISO}$ & $0.324 \pm 0.110$  \\ \\[-1em]
\hline
${r_{\rm p}, }_{\rm F18}$ & $0.514 \pm 0.517$  \\ \\[-1em]
${r_{\rm p}, }_{\rm B22}^{\rm LMC}$ & $0.300 \pm 0.252$  \\ \\[-1em]
${r_{\rm p}, }_{\rm B22}^{\rm L}$ & $0.246 \pm 0.250$  \\ \\[-1em]
${r_{\rm p}, }_{\rm B22}^{\rm H}$ & $0.281 \pm 0.229$  \\ \\[-1em]
${r_{\rm p}, }_{\rm P22}^{\rm LMC}$ & $0.318 \pm 0.245$  \\ \\[-1em]
${r_{\rm p}, }_{\rm P22}$ & $0.313 \pm 0.260$  \\ \\[-1em]
\hline
    \end{tabular}
    \caption{Typical uncertainties in the quantities used in this work. Col.~1 lists the set of $\rperi$ and $\rhocen$ being considered; col.~2 ($\left<\sigma_{\rm data}\right>$) gives the corresponding average uncertainty for that given quantity. }
    \label{tab:Scatter_lev}
\end{table}

In Table~\ref{tab:Scatter_lev} we show the average uncertainty ($\left<\sigma_{\rm data}\right>$) for each of the sets of $\rperi$ and $\rhocen$ analyzed, obtained from fitting the CDF of a log-normal distribution to the reported percentiles. To mimic these typical uncertainties, we then choose 
three different values of $\sigma_{\rm LN}=\{0.1,~0.25,~0.5\}$ (which hereafter we also refer to as 'noise levels' 1, 2, 3, respectively) with which to produce the mock data sets. For each combination of these $3$ noise levels, we explore $5$ random realizations, getting a total of $3\times3\times5=45$ pairs of mock data sets $\xi=\{\rhocen, \rperi\}$. 

\begin{figure}
    \centering
    \includegraphics[width=1\columnwidth]{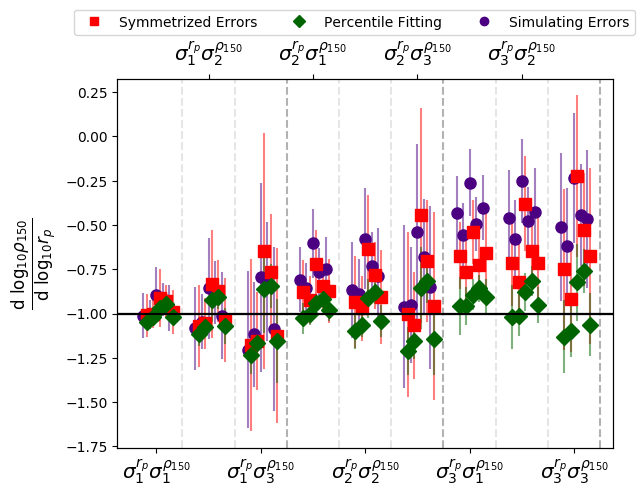}
    \caption{Recovered logarithmic slope for different mock data sets. The horizontal black line shows the input logarithmic slope ($-1$). The different symbols and colours indicate the methods tested (as indicated in the legend).  The groups of points refer to the noise levels explored, as indicated by the subscripts $1,~2,~3$ on $\sigma$.
    }
    \label{fig:MOC_Corr}
\end{figure}

Fig.~\ref{fig:MOC_Corr} shows the recovered logarithmic slope $m$ using the $3$ different fitting procedures applied to the mocks. This analysis suggests that both the method of \KAP{}, i.e. the 'symmetrized errors' method, and the 'simulating errors' method suffer from a bias towards flatter relations. Thus the slopes derived with these two methods will be considered as upper limits on the actual slope. The bias  worsens for larger noise levels and is mainly driven by the error level on the pericentric radii. The \KAP{} method yields slopes closer to the true values than the 'simulating errors' method; therefore, in the remainder of the article, we will not consider the 'simulating errors' method further. On the other hand, the percentile method appears to be the one with the best performance overall, yielding unbiased values of the slope. This will be our reference method.

\section{Results}
\label{sec:results}

\begin{figure*}
    \centering
    \includegraphics[width=1\textwidth]{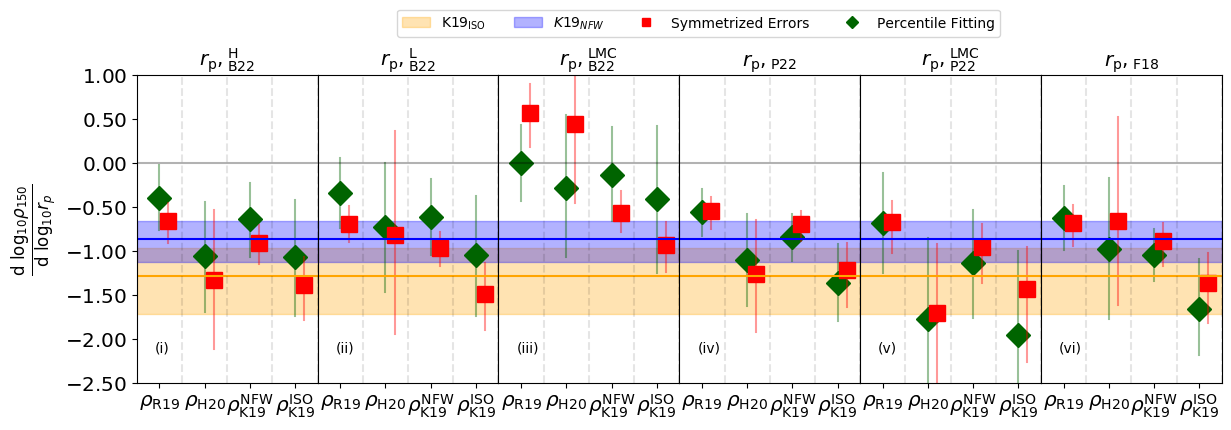}
     \caption{Measured logarithmic slope of the $\rhocen$-$\rperi$ relation for the different pairs of data sets. 
     The panels refer to the different sources of the measured pericentres: \BAT{}($i,~ii,~iii$);  \PAC{} ($iv,~v$) and  \FRI{} ($vi$). The source of the central density measurement is indicated in the horizontal axis of each panel. 
     Blue and orange lines indicate the reported values for the logarithmic slope in \KAP{} using \FRI{} pericentres, for an isothermal sphere (yellow) and a NFW profile (blue).
     Different symbols indicate different fitting method as indicated in the legend. 
    }
    \label{fig:Slopes}
\end{figure*}

Figure~\ref{fig:Slopes} shows the logarithmic slope recovered with the 'percentile fitting' and 'symmetrized errors' methods for all the data set pairs considered.
In general, the logarithmic slopes cover the range between $-0.5$ down to $-1.5$, hinting at an underlying anti-correlation, in agreement with previous studies. We also note that we recover the \KAP{} results, when using the same set of densities and pericentric radii as in their work (we recover them exactly when applying their same methodology, and well within $1\sigma$ when using the percentile fitting method, see last panel of Fig.~\ref{fig:Slopes}). The slope obtained with the \HAY{} densities are the most uncertain, due to the larger errors on $\rhocen$, most likely due to the more complex modelling performed by the authors, which allows from deviations from spherical symmetry in both the stellar and dark matter component of the dwarf galaxy. 

Since the percentile fitting method appears to be the most robust, in Fig.~\ref{fig:correlations} we show the best fits obtained using this method as a reference. In each panel we show the $68$ per cent confidence region of the power law fit as a shaded band. The colours of the shaded bands indicate whether the logarithmic slope $m$ is compatible with $0$ within $<1\sigma$ (grey), between $1\sigma$ and $3\sigma$ (blue) or more than $>3\sigma$ (magenta). 
As previously discussed, the value of $m$ is negative in most cases, but with a varying statistical significance depending on the combination of data sets considered. In general, the relation obtained using the \BAT{} $\rperi$ in the potential including the LMC are flatter than in the other cases 
with a logarithmic slope consistent with zero within $1\sigma$. 

When using the same pair of data sets as \KAP{}, we recover their result of a statistically significant  (at $\gtrsim 3 \sigma$) anti-correlation (see panels in the 2nd and 3rd row and rightmost column of Fig.~\ref{fig:correlations}). 
The only other combination of data sets for which we find that $m$ differs from zero in a statistically significant way is when the \KAP{} $\rhocen$ are paired with the \PAC{} $\rperi$. Apart from variations in $\rperi$, a contributing factor is that the \KAP{} densities for the galaxies with the smallest and largest pericentres tend to be, respectively, higher and lower than the corresponding estimates obtained in other works and also have smaller error bars. 

In order to test the effect of the small sample size in the derivation of the significance of the logarithmic slope, we repeated the analysis removing one of the galaxies at a time.
On average, the effect of removing one galaxy (from any of the data-sets) is to diminish the significance of the recovered logarithmic slope by a factor of $0.66$.
Furthermore, we do not find any galaxy whose removal systematically increases the significance of the relation or that would make
it become statistically significant if the original significance
is below $3\sigma$.

As a further investigation of what relation might be underlying the inferred $\rperi$ and $\rhocen$, we wish to compare models ($f_{\theta}$) with a different dependence of $\rhocen$ on $\rperi$. Specifically, we test two models in which $\rhocen$ depends on $\rperi$ and one in which it is independent: 
\begin{enumerate}
    \item[\textbf{Power Law~(PL)}] As in \KAP{}, the model is defined as $f_{\theta}\equiv f(r ~|~m, q) = 10^{q}r^{m}$, and the scatter $\sigma_{i}^{2} = {\delta \rho_{i}}^{2} + \left(m10^{q}r_{i}^{m-1}\delta r_{i}\right)^2$. \\
    \item[\textbf{Exponential~(EX)}] The model is defined as $f_{\theta}\equiv f(r~|~m, q) = \exp\left[q+ m r\right]$, and the scatter $\sigma_{i}^{2} = {\delta \rho_{i}}^{2} + \left(m\exp\left[q + m r_{i}\right]\delta r_{i}\right)^2$. \\
    \item[\textbf{Intrinsic Scatter~(IS)}]  The central density is independent of the pericentric distances. The model is defined as $f_{\theta}\equiv f(r ~|~m, \sigma_{0}) = m$, and the scatter $\sigma_{i}^{2} = {\delta \rho_{i}}^{2}+\sigma_{0}^2$, with $\sigma_{0}$ the intrinsic scatter. 
\end{enumerate}

In the formulae described above, we use the same normalization as in Eq.~\eqref{eq:PL}, with the DM density $\rho$ (and its uncertainty $\delta\rho$) normalized to $10^7~{\rm M_{\odot}\,kpc^{-3}}$, while the pericentric distance $r$ (and its uncertainty $\delta\rho$) is normalized to $1 {\rm kpc}$. 

We compare the quality of each of the previous models by making use of the Akaike Information Criterion \citep[AIC ; ][]{Akaike1974} including a correction due to small sample size \citep{burnham1998model}:
\begin{equation}
    {\rm AIC}_{c} = 2k - 2\log{\mathcal{L}^{\rm max}} + \frac{2k(k+1)}{n-k-1}, 
\end{equation}
with $k$ the number of free parameters of the model tested. The model comparison is done using the metric $\Delta_{i} = {\rm AIC}_{c}^{i} - {\rm min}_{j}~{\rm AIC}_{c}^{j}$. Those models with $\Delta_{i}<2$ have substantial empirical support and cannot be rejected  \protect\citep{burnham1998model}.

Since calculating a likelihood for the percentile fitting method is a rather complex business, we take advantage of the fact that the percentile fitting and the symmetrized error methods give comparable results to perform the model comparison via the latter method, which allows for a simpler Gaussian likelihood; then, we define the likelihood as:
\begin{equation}
    \mathcal{L}\left(\{r_{i}\}, \{\rho_{i}\} | \theta\right) = \prod_{i} \frac{1}{\sqrt{2\pi\sigma^{2}_{i}}}\exp{\left[-\frac{\left(\rho_i - f_{\theta}(r_i)\right)^2}{2\sigma^{2}_{i}}\right]} , 
\end{equation}
with $\sigma_{i}^{2}$ the variance, $f_{\theta}$ the model, and $\theta$ a vector gathering the model parameters. For sampling the likelihood we have performed a MCMC analysis as implemented in the public python package \textit{emcee} \citep[][]{emcee}. 

A summary of the model comparison can be found in Tab.~\ref{tab:AIC} (see Tab.~\ref{tab_apx1:Models} for the parameters of the best-fitting models to the different pairs of data sets). 
We find that, for an isolated MW,  the two models with a decreasing central DM density of the dSphs as a function of $\rperi$ (PL and EX) are preferred and perform similarly well. This preference is approximately independent of the source of the central density estimates. The only relevant exception is the densities from \HAY{} for which models with or without a dependence on $\rperi$ have similar empirical support. Also, with this analysis the \KAP{} densities are those that provide the largest support to the models with decreasing $\rhocen$ as a function of $\rperi$. 

The effect of the inclusion of the LMC infall, on the orbital integration of the dSphs appears to yield discrepant results between the two works that consider it.
The pericentric distances derived by \BAT{} seem to prefer the model where $\rhocen$ and $\rperi$ are uncorrelated, 
while $3$ out of the $4$ data pairs using \PAC{} pericentres prefer the models in which $\rhocen$ depends on $\rperi$. 
This is due to differences between the two sets of determinations which, while within $1$ or at most $2\sigma$, do nevertheless move the points on the $\rhocen$ vs $\rperi$ plane enough to change the significance of the relation.  
Nonetheless, the empirical support for the models with a dependence on pericentric distance becomes much milder when including the effect of the LMC with respect to when considering an isolated potential (see Tab.~\ref{tab:AIC}), and also when using the \PAC{} determinations of $\rperi$. The inclusion of the LMC infall onto the MW seems to be an important ingredient for exploring the existence of this possible anti-correlation, and further efforts to understand the effect of the LMC on the orbital properties of the MW satellites are needed.

\begin{table}
    \centering
\begin{tabular}{lllccc}
\hline
$\rperi$ &          $\rhocen$ & Preferred Models  &  $\Delta_{\rm EX}$ &  $\Delta_{\rm PL}$ & $\Delta_{IS}$ \\
\hline\hline
   B22\_H &      R19 &             EX\textbackslash PL &  \textbf{0.0} &  \textbf{0.5} &  2.1 \\
   B22\_H &      H20 &             EX\textbackslash PL\textbackslash IS  &  \textbf{0.5} &  \textbf{0.0} &  \textbf{1.1} \\
   B22\_H &  K19\_NFW &             EX\textbackslash PL &  \textbf{0.9} &  \textbf{0.0} &  8.4 \\
   B22\_H &  K19\_ISO &             EX\textbackslash PL &  \textbf{0.0} &  \textbf{0.3} &  6.8 \\ \hline
   B22\_L &      R19 &             EX\textbackslash PL &  \textbf{0.0} &  \textbf{1.3} &  3.4 \\
   B22\_L &      H20 &             EX\textbackslash PL\textbackslash IS &  \textbf{1.4} &  \textbf{1.4} &  \textbf{0.0} \\
   B22\_L &  K19\_NFW &             EX\textbackslash PL &  \textbf{0.6} &  \textbf{0.0} &  9.1 \\
   B22\_L &  K19\_ISO &             EX\textbackslash PL &  \textbf{0.0} &  \textbf{0.8} &  7.3 \\ \hline
 B22\_LMC &      R19 &             \hspace{3em}IS &  4.1 &  5.3 &  \textbf{0.0} \\
 B22\_LMC &      H20 &             EX\textbackslash PL\textbackslash IS &  \textbf{0.8} &  \textbf{1.0} &  \textbf{0.0} \\
 B22\_LMC &  K19\_NFW &             \hspace{3em}IS &  8.7 & 11.9 &  \textbf{0.0} \\
 B22\_LMC &  K19\_ISO &             \hspace{3em}IS &  3.0 &  5.6 &  \textbf{0.0} \\ \hline
     P22 &      R19 &             EX\textbackslash PL &  \textbf{0.6} &  \textbf{0.0} &  6.1 \\
     P22 &      H20 &             EX\textbackslash PL &  \textbf{0.9} &  \textbf{0.0} &  2.6 \\
     P22 &  K19\_NFW &             EX\textbackslash PL &  \textbf{1.2} &  \textbf{0.0} & 10.5 \\
     P22 &  K19\_ISO &             EX\textbackslash PL &  \textbf{0.0} &  \textbf{0.1} &  9.0 \\ \hline
 P22\_LMC &      R19 &             EX\textbackslash PL\textbackslash IS &  \textbf{0.0} &  \textbf{0.7} &  \textbf{1.9} \\ 
 P22\_LMC &      H20 &             EX\textbackslash PL &  \textbf{0.1} &  \textbf{0.0} &  2.6 \\
 P22\_LMC &  K19\_NFW &             EX\textbackslash PL &  \textbf{0.0} &  \textbf{0.6} &  3.3 \\
 P22\_LMC &  K19\_ISO &             EX\textbackslash PL &  \textbf{0.0} &  \textbf{0.8} &  4.2 \\ \hline
     F18 &      R19 &             EX\textbackslash PL &  \textbf{0.4} &  \textbf{0.0} &  3.5 \\
     F18 &      H20 &             EX\textbackslash PL\textbackslash IS &  \textbf{1.2} &  \textbf{1.0} &  \textbf{0.0} \\
     F18 &  K19\_NFW &             EX\textbackslash PL &  \textbf{0.5} &  \textbf{0.0} &  7.3 \\
     F18 &  K19\_ISO &             EX\textbackslash PL &  \textbf{0.0} &  \textbf{0.6} &  5.9 \\
\hline
\hline
\end{tabular}
    \caption{Results of model comparison for the different data sets
    pairs. The model comparison is done using the metric $\Delta_{i} = {\rm AIC}_{c}^{i} - {\rm min}_{j}~{\rm AIC}_{c}^{j}$. Those models with $\Delta_{i}<2$ have substantial empirical support and cannot be rejected  \protect\citep{burnham1998model}. We explicitly identify them in the column 'Preferred Models' and their $\Delta_{i}$ is highlighted in boldface in the following columns.  
    }
    \label{tab:AIC}
\end{table}

\section{Discussion and Summary} \label{sec:conclusions}

In this contribution we re-assess the question of whether an anti-correlation exists between the central DM density ($\rhocen$) and orbital pericentre ($\rperi$) of classical dSph satellites of the MW. We explore in a quantitative way how the existence and strength of the relation depend on the adopted sets of inferred central DM densities and pericentric distances. Specifically, we consider the $\rhocen$ determinations by \READ{}, \KAP{} and \HAY{}, and the pericentric radii by \BAT{} and \PAC{} in different gravitational potentials for the MW, both isolated or including the infall of a massive LMC. We also test a new method designed to handle the asymmetric error bars that naturally appear during the measurement of non-negative quantities such as radial distances or densities; this method is of general applicability to all situations in which asymmetric errors are present. 

In general, we find that only the adoption of the \KAP{} DM densities with the $\rperi$ from the earlier Gaia DR2 determinations or the isolated MW potential by \PAC{} results in strong empirical support for models in which $\rhocen$ and $\rperi$ are anti-correlated, and a logarithmic slope differing from zero at $>3 \sigma$ when exploring a power-law dependence. In contrast, the \HAY{} DM densities do not result in a preference for models in which $\rhocen$ and $\rperi$ are anti-correlated with respect to models where they are independent. Clearly, uncertainties in the determinations of the $\rhocen$ play a role in establishing whether the existence of such anti-correlation is robust. The same can be said for the determinations of the pericentric radii. Even though the 3D bulk motions of the MW classical dSphs are now known in exquisite detail, transforming these observables into pericentric radii requires the adoption of a gravitational potential in which to integrate the orbits, which results in additional uncertainties. Besides the fact that the mass of the MW is still unknown within a factor of $2$ \citep[see e.g.][]{Fri20, Wan22}, it is clear that the inclusion of the infall of a massive LMC in the determinations of the orbits pf MW dSphs impacts the resulting $\rperi$, and that there is some variation in the values determined for these $\rperi$ across different studies.

The issue is exacerbated by the small number of galaxies involved in the analysis, which implies that any conclusion should be taken with care. 
Works by \KAP{} and \cite{Hay22} suggest that increasing the sample through the inclusion of MW UFD galaxies completely washes away the anti-correlation present in the MW dSphs. 
It is not clear whether the reason is to be ascribed to the larger errors in the inferences of the DM densities and pericentric radii of UFDs or to some more fundamental property of these galaxies. 
For example, a fraction of the UFDs may be on their first infall onto the MW as can be gathered by results of orbital integration (e.g. \BAT{}, \PAC{}) and arguments on conservation of energy and angular momentum (e.g. \citealt{Hammer2021}). Thus, those galaxies may not have suffered from the tidal field of the MW, and consequently can present small pericenter distances abut central densities unaffected by their orbits, thus obscuring the possible relation. 
We plan to explore the dependence of this anti-correlation on the inclusion of UFDs in future work

The aforementioned possible effect of first infall is not necessarily limited to UFDs. 
The orbital integration of LeoI indicates that this classical dSph is on its first infall, having just passed its first pericenter.  Furthermore, close interaction with the LMC may also introduce outliers in the relation.  
In particular, Carina has a small but non-negligible probability of being related to the LMC (\BAT{}, \PAC{}). 
However, as discussed in Sec.~\ref{sec:results}, the significance of the obtained relations is not driven by any specific subset of satellites, indicating that our results are robust. 

Not only is the existence of the anti-correlation controversial, but also its origin. A caveat against the survival bias hypothesis is the lack of MW satellites with high inner DM densities on external orbits, i.e. with large pericentres. This problem is reduced when adopting the pericentres of \BAT~(with the LMC): in this case Draco and Leo~I populate the high-$\rhocen$ and high-$\rperi$ region of the parameter space.

The results from simulations are also contradictory. On the one hand, satellites in DM-only simulations seem to show an anti-correlation between the central dark matter density and the distance of the most recent pericentric passage (see \HAY{}). On the other hand, the inclusion of baryons in the simulations may have a strong impact on the proposed relation. As shown in \citet[][]{Robles2019}, the inclusion of the potential of the MW disc in the simulations can strongly alter the density of the satellites, diminishing the DM densities of the satellites with smaller pericentres, and inverting the relation. 

In summary, out of the $24$ combinations of $\rperi$ and $\rhocen$ explored, we found that only $3$ strongly support (at more than $3\sigma$ level) the presence of an anti-correlation between those two quantities: these represent the $12.5$ per cent of the models explored. When making use of the AIC for model comparison, we find the data to be better described by models in which the central density $\rhocen$ decreases as a function of $\rperi$ (power-law and exponential), and these perform similarly well. Only in one of the MW potentials explored is the data better described by a model with no dependence between $\rhocen$ on $\rperi$. 
Our results suggest that the strength and the existence of the $\rhocen$-$\rperi$ anti-correlation on the MW's dSphs is still debatable.
Exploring the existence and characteristics of this relationship with cosmological simulations at very high resolution, including properly modelled baryonic effects, as well as different DM flavours, will be a step forward towards understanding its emergence and its likelihood in a $\Lambda$CDM universe. We plan to do this in a future contribution.

\section*{Acknowledgements}
S.C.B. acknowledges support from the Spanish Ministry of Economy and Competitiveness (MINECO) under the grant SEV-2015-0548-18-3 and the Spanish Ministry of Science and Innovation (MICIU/FEDER) through research grant PGC2018-094975-C22. G.B., S.C.B. acknowledge support from the Agencia Estatal de Investigación del Ministerio de Ciencia en Innovación (AEI-MICIN) and the European Regional Development Fund (ERDF) under grant number PID2020-118778GB-I00/10.13039/501100011033. G.B. acknowledges the AEI under grant number CEX2019-000920-S. A.D.C. is supported by a Junior Leader fellowship from `La Caixa' Foundation (ID 100010434), fellowship code  LCF/BQ/PR20/11770010. 

Data analysis was performed using the Python\footnote{\url{https://www.python.org}} programming language. The following Python modules were used for the analysis: pandas \citep{pandas}; numpy \citep{numpy}; scipy \citep{SciPy}; matplotlib \citep{matplotlib}; corner \citep{corner}; numba \citep{numba}; h5py \citep{h5py} and emcee \citep{emcee}.

\section*{Data Availability}

The data underlying this article will be shared on reasonable request to the corresponding author.

\bibliography{biblio_rhorp}
\bibliographystyle{mnras}

\appendix

\section{Simulated errors}
\label{Apx:Resimulation}

Our first approach for handling asymmetric errors is to describe the error distribution of each data point with some probability distribution. Once this distribution has been chosen, we sample new data from it and fit them.

The procedure can be summarised as follows: 
\begin{enumerate}
    \item\label{item:ChoosePD} Choose a probability distribution to be fit to the percentiles $x_{50\mathrm{th}},~x_{16\mathrm{th}},~x_{84\mathrm{th}}$;
    \item\label{item:FitPD} For each galaxy ($g$) and variable $\xi^{g} = \{\rperi^{g}, \rhocen^{g}\}$, obtain the best fitting parameters of the CDF corresponding to item (i);
    \item\label{item:Sample} Randomly sample for each galaxy one value of the density $\rhocen^{g, i}$ and pericentric distance $\rperi^{g, i}$ following the distribution obtained in \ref{item:FitPD}. Lets denote these new data as $\xi^{g, i}$;
    \item\label{item:FitRelation} Fit the desired relation between re-sampled densities and pericentres, obtaining this way the set of parameters $\theta^{i}$;
    \item Repeat \ref{item:Sample} and \ref{item:FitRelation} $N$ times: ${\boldsymbol \theta} = \{\theta^{1}, \theta^{2}, ..., \theta^{N}\}$, with $N=10^4$; 
    \item Store the results as the median of ${\boldsymbol \theta}$. 
\end{enumerate}

For point (i), we tested three different distributions: a log-normal (LN) distribution, a log-logistic (LL) distribution and a Gaussian  distribution truncated at zero (TG). Since we find that these give similar results, in the main text we only show those for the LN distribution, and give the corresponding equation for the PDF, $f_{\rm LN}$, and CDF, $F_{\rm LN}$, below:

\begin{align}
\label{eq:logNorm}
    f_{\rm LN}(x~|~\mu_{\rm LN}, \sigma_{\rm LN}) &= \frac{1}{x\sigma_{\rm LN}\sqrt{2\pi}}\exp\left[{-\frac{\log x - \log \mu_{\rm LN}}{2\sigma_{\rm LN}^{2}}}\right],\\ \nonumber
    F_{\rm LN}(x~|~\mu_{\rm LN}, \sigma_{\rm LN}) &= \frac{1}{2}\left[1 + {\rm erf}\left(\frac{\log x  - \log \mu_{\rm LN}  }{\sqrt{2} \sigma_{\rm LN}}\right)\right]. \\
\end{align}

\section{Percentiles as random variables}
\label{Apx:ThreeOrderStat}

Our second approach for handling asymmetric errors is to treat each percentile as a random variable and model its probability distribution via order statistics.

Let us denote the available sets of $\rhocen$ and $\rperi$ for each galaxy as $\mathbf{d} =\{\rhocen,~\rperi \}$. In general, these data will be different from the "real" values, which we will denote as $\real{\mathbf{d}}=\{\real{\rhocen},~\real{\rperi}\}$. We indicate with $\mathbf{\eta}$ all the extra parameters of the model. Our goal is to obtain the probability distribution of the parameters $\{\real{\mathbf{d}}, \eta\}$ given our data $\mathbf{d}$, i.e. $P(\real{\mathbf{d}}, \eta~|~\mathbf{d}) $. 

We can decompose the probability described above via the Bayes' theorem as
\begin{equation}
\label{eq:posterior}
    P(\real{\mathbf{d}}, \eta~|~\mathbf{d})  = 
    \frac{1}{P(\mathbf{d})}P(\eta)P(\mathbf{d}, \real{\mathbf{d}}~|~\eta).
\end{equation}
The term in the denominator is a normalization constant that we can obviate. The second term in the r.h.s.\ is the a priori probability distribution of the parameters $\eta$. The last term is the joint probability distribution of the measured data $\mathbf{d}$ and the "real" values $\real{\mathbf{d}}$. We can further decompose this last term as
\begin{equation}
\label{eq:psudo-likelihood}
    P(\mathbf{d}, \real{\mathbf{d}}~|~\eta) = P(\mathbf{d}~|~\real{\mathbf{d}}, \eta)P(\real{\mathbf{d}}~|~\eta)
\end{equation}

The probability of the data given the "real" values $ P(\mathbf{d}~|~\real{\mathbf{d}}, \eta)$ and the probability distribution of the "real" values given the model we are testing $P(\real{\mathbf{d}}~|~\eta)$ remains to be solved for.

Let us start with the model. The second term in the r.h.s. of Eq.~\ref{eq:psudo-likelihood} can be decomposed as: $P(\real{\mathbf{d}}~|~\eta) = P(\real{\rhocen}~| ~\real{\rperi}, \eta)P(\real{\rperi}~| ~\eta)$. The first term is the conditional probability of the central DM density of the dSphs given the pericentric distance. The second term is the a priori probability distribution of the pericentres, which we will assume uniform between $r_{\rm min}$ and $r_{\rm max}$: $P(\real{\rperi}~|~\eta)= U_{\left[r_{\rm min}, r_{\rm max}\right]}(\real{\rperi})$. We model the relation between the central densities and the pericentres  as Eq.~(\ref{eq:PL}); the conditional probability of $\rhocen$ given $\rperi$ can be written as  
\begin{align}
    P(\real{\rhocen}~&|~\real{\rperi}, \eta) = \mathcal{N}\left(\log_{10}(\real{\rhocen})~\bigr|~q + m\log_{10}(\real{\rperi})~;~\sigma^{2}\right) 
    = \\\nonumber
    \frac{1}{\sqrt{2\pi\sigma^{2}}}&\exp\left\{-\frac{1}{{2\sigma^2}}\left[\log_{10}\left(\frac{\rhocen^{\ast}}{10^7~{\rm M_{\odot}/kpc^{3}}}\right) - \left(q + m\log_{10}\left.\frac{\rperi^{\ast}}{\rm kpc}\right)\right)\right]^{2}\right\},
\end{align}

The only term that remains to be solved for in Eq.~\ref{eq:psudo-likelihood} is the conditional probability
 of the measured data $\mathbf{d}$ given the "real" ones $\real{\mathbf{d}}$ i.e. $P(\mathbf{d}~|~\real{\mathbf{d}}, \eta)$. To characterize this last term we need to give more details on the data gathering process. First we make the reasonable assumption that the measurements of the central density and the pericentreic distance are independent. Thus we can split the probability distribution as: 
\begin{equation}
P(\mathbf{d}~|~\real{\mathbf{d}}, \eta) = P(\rhocen~|~\real{\rhocen}, \eta)P(\rperi~|~\real{\rperi}, \eta)~.
\end{equation}

The available estimates of $\rhocen$ and $\rperi$  are in the form of percentiles, namely $x_{16\mathrm{th}}$, $x_{50\mathrm{th}}$, $x_{84\mathrm{th}}$, where $x$ is either $\rhocen$ or $\rperi$. We can handle these data without losing information via order statistics. 

Let us assume a continuous PDF for $X$, $f(x; \beta)$ defined by the set of parameters $\beta$. If we get $N$ samples from the distribution $f$ and we sort them in such a way that: $x_1 < x_2 < \dots < x_{N-1} < x_{N}$, the joint probability distribution of all the order statistics described previously \citep{Arnold2008} is

\begin{equation}
    f_{X_{1}, ..., X_{N}}(x_{1},..., x_{N}) = N!\prod_{i=1}^{N} f(x_{i}; \beta).
\end{equation}

As we have only $3$ of the $N$ order statistics ($x_{i} < x_{k} < x_{j}$), we shall marginalize over the unwanted $N-3$ parameters. The result of this procedure is the joint probability distribution of three order statistics:

\begin{align}
    P\bigr(x_i, x_k, x_j &~\bigr|~ \beta, N\bigr) =  \frac{N!}{(i-1)!(N-j)!(k-i-1)!(j-k-1)!}~ \\ \nonumber
    \times~ & f\bigr(x_i~\bigr|~\beta\bigr)~f\bigr(x_k~\bigr|~ \beta\bigr)~f\bigr(x_j~\bigr|~\beta\bigr)~
    F\bigr(x_i~\bigr|~\beta\bigr)^{i-1}~\bigr(1-F\bigr(x_j~\bigr|~\beta\bigr)\bigr)^{N-j} ~ \\ \nonumber
    \times~ &\bigr(F\bigr(x_k~\bigr|~ \beta\bigr)-F\bigr(x_i~\bigr|~ \beta\bigr)\bigr)^{k-i-1}\bigr(F\bigr(x_j~\bigr|~ \beta\bigr) - F\bigr(x_k~\bigr|~ \beta\bigr)\bigr)^{j-k-1}.  
\end{align}
The indices $\{i, k, j\}$ of the order statistics can be easily related to the $p_{\{i, k, j\}}-\mathrm{th}$ percentile as\footnote{This is only true if the fraction is an integer. If not, the percentile is usually computed as the the weighted mean between the closest order statistics. Taking into account this fact is possible but increases the complexity of the model. Moreover, the difference is only significant when the number of samples $N$ is small. As this is not generally the case, we decided to stick with the simpler model.} $\{i, k, j\} = {\rm int}\left[1+p_{\{i, k, j\}}(N-1)/100\right]$. Note that this probability depends on the family chosen for the PDF (CDF) $f$($F$), the set of parameters $\beta$ and on the number of samples $N$. In particular, in this work $\beta=\{\real{x}, s^{2}\}$ where $\real{x}$ is a location parameter while $s^{2}$ a parameter characterizing the width of the distribution. In particular in this work we chose a log-normal PDF parameterized by the median $\real{x}$ and the variance in log space $s^2$: 
\begin{equation}
    f(x~|~\beta=\{\real{x}, s^2\} ) = \frac{1}{x\sqrt{2\pi s^2}} \exp{\left[-\frac{
    \ln^{2}{x/\real{x}}}{2 s^2}\right]}.
\end{equation}
If we introduce a prior over $N$ and $s^{2}$ we can remove these two parameters via marginalization. We have assumed a uniform prior for $N$, between $N_{min}=20$ and $N_{max}=10^4$ such as $P(N) = U_{[N_{\rm min}, N_{\rm max})}(N)$. For $s^{2}$ we impose as prior a scaled inverse chi square distribution:
\begin{equation}
    P_{{\rm Scale-Inv-}\chi^2}(s^{2}~|~\nu, \tau^{2}) = \frac{(\tau^{2}\nu/2)^{\nu/2}}{\Gamma(\nu/2)}\frac{\exp\left[-\displaystyle \frac{\tau^{2}\nu}{2s^{2}}\right]}{(s^{2})^{1 + \nu/2}},
\end{equation}
with one degree of freedom ($\nu = 1$) and fixed scale parameter ($\tau^{2}=0.1$), with $\Gamma$ being the gamma function.

Finally the probability distribution of the measured data given the "real" values $P(\mathbf{d}| \real{d}, \eta)$ can be written as
\begin{align}
\label{eq:observed-given-measured}
    P(\mathbf{d}| \real{d}, \eta) & = \prod_{\displaystyle\xi=\{\rhocen,\rperi\}} \int{\rm d}{s^{2}}\sum_{N} \biggr[
    U_{[N_{\rm min}, N_{\rm max})}(N) \\ \nonumber
    \times ~ & P_{{\rm Scale-Inv-}\chi^2}(s^{2}~|~\nu=1, \tau^{2}=0.1) \\ \nonumber
    \times ~ & P\bigr(\xi_i, \xi_k, \xi_j ~\bigr|~ \beta=\{\real{\xi}, s^{2}\}, N\bigr)
     \biggr].
\end{align}

The parameters we are finally left to fit are $\{\real{\rperi}\}$, $\{\real{\rhocen}\}$, and ${\eta=\{m, q, \sigma^{2}\}}$, i.e.\ $2n+3$ parameters with $n=8$ the number of galaxies. We also need to choose the prior probability over $\eta$. For $m$ and $q$ we choose uniform probability distributions $P(m) = U_{(m_{\rm min}, m_{\rm max})}(m)$ and $P(q) = U_{(q_{\rm min}, q_{\rm max})}(q)$. For $\sigma^2$ we chose a prior of the form $P(\sigma^2) \propto 1/\sigma^2$. The values used in this work can be found in Tab.~\ref{tab:fix_params}.

\begin{table}
    \centering
    \begin{tabular}{c|c}
    \hline
        Parameter & Value\\ \hline\hline
        $N_{\rm min}$, $N_{\rm max}$ & $10$, $10^4$  \\
        $r_{\rm min}$, $r_{\rm max}$ & $10^{-3}$, $5\cdot10^2$  \\
        $m_{\rm min}$, $m_{\rm max}$ & $-15$, $15$  \\
        $q_{\rm min}$, $q_{\rm max}$ & $-15$, $15$  \\
        $\nu$, $\tau^{2}$            & $1$, $0.1$  \\
        \hline\hline
    \end{tabular}
    \caption{Fixed parameters of the model \ref{eq:posterior}.}
    \label{tab:fix_params}
\end{table}

\section{Model comparison}
\label{Apx:comparison}

We report here Table~\ref{tab_apx1:Models}, containing the results of the MCMC sampling of the likelihood described in Sect.~\ref{sec:results}. 


\begin{table*}
    \centering
\begin{tabular}{llcccccc}
\hline
    &       &   \multicolumn{2}{c}{Intrinsic scatter} & \multicolumn{2}{c}{Power Law} & \multicolumn{2}{c}{Exponential}   \\
$\rperi$ &  $\rhocen$ & $m$ & $\sigma_{0}^{2}$ & $q$ & $m$ & $q$ & $m$ \\
\hline\hline
B22\_H  &  R19  &  ${15.3}_{-2.2}^{+2.2}$  &  ${30}_{-16}^{+41}$  &  ${2.3}_{-0.3}^{+0.4}$  &  ${-0.9}_{-0.2}^{+0.2}$  &  ${3.4}_{-0.2}^{+0.3}$  &  ${-0.012}_{-0.005}^{+0.004}$\\ \\[-1em]
B22\_H  &  H20  &  ${17.3}_{-4.7}^{+6.0}$  &  ${111}_{-73}^{+189}$  &  ${3.4}_{-1.4}^{+1.3}$  &  ${-1.3}_{-0.8}^{+0.8}$  &  ${4.1}_{-1.0}^{+0.8}$  &  ${-0.027}_{-0.016}^{+0.017}$\\ \\[-1em]
B22\_H  &  K19\_NFW  &  ${15.3}_{-2.5}^{+2.7}$  &  ${45}_{-23}^{+57}$  &  ${2.8}_{-0.3}^{+0.4}$  &  ${-0.9}_{-0.2}^{+0.2}$  &  ${3.7}_{-0.2}^{+0.3}$  &  ${-0.018}_{-0.005}^{+0.005}$\\ \\[-1em]
B22\_H  &  K19\_ISO  &  ${13.1}_{-3.2}^{+3.6}$  &  ${71.5}_{-38}^{+94.4}$  &  ${3.5}_{-0.6}^{+0.7}$  &  ${-1.4}_{-0.4}^{+0.3}$  &  ${4.0}_{-0.4}^{+0.4}$  &  ${-0.027}_{-0.009}^{+0.007}$\\  \hline \\[-1em]
B22\_L  &  R19  &  ${15.3}_{-2.2}^{+2.2}$  &  ${29.9}_{-16.4}^{+41.3}$  &  ${2.5}_{-0.4}^{+0.4}$  &  ${-0.7}_{-0.2}^{+0.2}$  &  ${3.4}_{-0.2}^{+0.2}$  &  ${-0.009}_{-0.003}^{+0.003}$\\ \\[-1em]
B22\_L  &  H20  &  ${17.2}_{-4.8}^{+5.9}$  &  ${110}_{-72.7}^{+183}$  &  ${2.6}_{-2.2}^{+2.1}$  &  ${-0.8}_{-1.1}^{+1.2}$  &  ${3.2}_{-1.2}^{+1.4}$  &  ${-0.010}_{-0.020}^{+0.015}$\\ \\[-1em]
B22\_L  &  K19\_NFW  &  ${15.3}_{-2.5}^{+2.6}$  &  ${45}_{-23}^{+59}$  &  ${3.0}_{-0.3}^{+0.4}$  &  ${-1.0}_{-0.2}^{+0.2}$  &  ${3.6}_{-0.3}^{+0.2}$  &  ${-0.012}_{-0.003}^{+0.003}$\\ \\[-1em]
B22\_L &  K19\_ISO  &  ${13.1}_{-3.2}^{+3.5}$  &  ${71}_{-38}^{+96}$  &  ${3.9}_{-0.6}^{+0.7}$  &  ${-1.5}_{-0.4}^{+0.4}$  &  ${4.0}_{-0.4}^{+0.4}$  &  ${-0.020}_{-0.006}^{+0.005}$\\ \hline \\[-1em]
B22\_LMC  &  R19  &  ${15.3}_{-2.2}^{+2.3}$  &  ${30}_{-17}^{+42}$  &  ${0.1}_{-0.6}^{+0.7}$  &  ${0.6}_{-0.4}^{+0.3}$  &  ${2.2}_{-0.3}^{+0.8}$  &  ${+0.006}_{-0.011}^{+0.004}$\\ \\[-1em]
B22\_LMC  &  H20  &  ${17.2}_{-4.7}^{+6.0}$  &  ${111}_{-73}^{+187}$  &  ${0.2}_{-1.6}^{+1.8}$  &  ${0.4}_{-0.9}^{+0.8}$  &  ${2.0}_{-0.8}^{+0.9}$  &  ${+0.005}_{-0.010}^{+0.008}$\\ \\[-1em]
B22\_LMC  &  K19\_NFW  &  ${15.3}_{-2.6}^{+2.7}$  &  ${45}_{-23}^{+57}$  &  ${2.2}_{-0.5}^{+0.4}$  &  ${-0.6}_{-0.2}^{+0.2}$  &  ${3.3}_{-0.3}^{+0.2}$  &  ${-0.008}_{-0.003}^{+0.003}$\\ \\[-1em]
B22\_LMC  &  K19\_ISO  &  ${13.1}_{-3.3}^{+3.6}$  &  ${72}_{-38}^{+96}$  &  ${2.8}_{-0.5}^{+0.6}$  &  ${-0.9}_{-0.3}^{+0.3}$  &  ${3.4}_{-0.3}^{+0.3}$  &  ${-0.013}_{-0.004}^{+0.003}$\\ \hline \\[-1em]
P22  &  R19  &  ${15.3}_{-2.2}^{+2.2}$  &  ${30}_{-16}^{+41}$  &  ${2.2}_{-0.3}^{+0.4}$  &  ${-0.6}_{-0.2}^{+0.2}$  &  ${3.2}_{-0.2}^{+0.2}$  &  ${-0.008}_{-0.003}^{+0.002}$\\ \\[-1em]
P22  &  H20  &  ${17.2}_{-4.7}^{+5.9}$  &  ${111}_{-73}^{+191}$  &  ${3.5}_{-1.1}^{+1.2}$  &  ${-1.3}_{-0.7}^{+0.6}$  &  ${4.1}_{-0.8}^{+0.8}$  &  ${-0.022}_{-0.012}^{+0.012}$\\ \\[-1em]
P22  &  K19\_NFW  &  ${15.3}_{-2.5}^{+2.6}$  &  ${44}_{-23}^{+56}$  &  ${2.5}_{-0.3}^{+0.3}$  &  ${-0.7}_{-0.2}^{+0.2}$  &  ${3.5}_{-0.2}^{+0.2}$  &  ${-0.011}_{-0.003}^{+0.003}$\\ \\[-1em]
P22  &  K19\_ISO  &  ${13.1}_{-3.3}^{+3.6}$  &  ${71}_{-38}^{+94}$  &  ${3.3}_{-0.6}^{+0.7}$  &  ${-1.2}_{-0.4}^{+0.3}$  &  ${3.7}_{-0.3}^{+0.4}$  &  ${-0.017}_{-0.007}^{+0.005}$\\ \hline \\[-1em]
P22\_LMC  &  R19  &  ${15.3}_{-2.2}^{+2.7}$  &  ${30}_{-17}^{+41}$  &  ${2.4}_{-0.4}^{+0.6}$  &  ${-0.7}_{-0.4}^{+0.2}$  &  ${3.3}_{-0.2}^{+0.3}$  &  ${-0.010}_{-0.005}^{+0.004}$\\ \\[-1em]
P22\_LMC  &  H20  &  ${17.4}_{-4.8}^{+5.8}$  &  ${110}_{-73}^{+186}$  &  ${4.2}_{-1.5}^{+1.6}$  &  ${-1.7}_{-0.9}^{+0.8}$  &  ${4.5}_{-0.9}^{+0.9}$  &  ${-0.029}_{-0.015}^{+0.013}$\\ \\[-1em]
P22\_LMC  &  K19\_NFW  &  ${15.3}_{-2.6}^{+2.6}$  &  ${45}_{-23}^{+56}$  &  ${2.9}_{-0.5}^{+0.7}$  &  ${-1.0}_{-0.4}^{+0.3}$  &  ${3.6}_{-0.2}^{+0.3}$  &  ${-0.014}_{-0.006}^{+0.004}$\\ \\[-1em]
P22\_LMC  &  K19\_ISO  &  ${13.1}_{-3.3}^{+3.6}$  &  ${73}_{-40}^{+94}$  &  ${3.6}_{-0.8}^{+1.4}$  &  ${-1.4}_{-0.8}^{+0.5}$  &  ${3.8}_{-0.4}^{+0.7}$  &  ${-0.020}_{-0.013}^{+0.007}$\\ \hline \\[-1em]
F18  &  R19  &  ${15.3}_{-2.2}^{+2.2}$  &  ${29}_{-16}^{+40}$  &  ${2.4}_{-0.4}^{+0.5}$  &  ${-0.7}_{-0.3}^{+0.2}$  &  ${3.4}_{-0.2}^{+0.3}$  &  ${-0.010}_{-0.004}^{+0.003}$\\ \\[-1em]
F18  &  H20  &  ${17.4}_{-4.8}^{+5.9}$  &  ${111}_{-73}^{+190}$  &  ${2.4}_{-2.3}^{+1.7}$  &  ${-0.7}_{-0.9}^{+1.2}$  &  ${3.4}_{-1.4}^{+1.2}$  &  ${-0.012}_{-0.015}^{+0.016}$\\ \\[-1em]
F18  &  K19\_NFW  &  ${15.2}_{-2.5}^{+2.7}$  &  ${45}_{-23}^{+58}$  &  ${2.8}_{-0.4}^{+0.5}$  &  ${-0.9}_{-0.3}^{+0.2}$  &  ${3.7}_{-0.2}^{+0.3}$  &  ${-0.014}_{-0.004}^{+0.003}$\\ \\[-1em]
F18  &  K19\_ISO  &  ${13.1}_{-3.3}^{+3.6}$  &  ${72}_{-39}^{+94}$  &  ${3.6}_{-0.6}^{+0.9}$  &  ${-1.4}_{-0.5}^{+0.3}$  &  ${3.9}_{-0.4}^{+0.4}$  &  ${-0.020}_{-0.006}^{+0.005}$\\ \\[-1em]
\hline
\hline
\end{tabular}
    \caption{
    Results of the MCMC sampling of the likelihood described in Sect.~\ref{sec:results}. The first two columns indicate the data-set pair used in the regression.  The following columns indicate the median of the parameters of the three different models tested in this work. The lower and upper errorbars indicate the $16^\mathrm{th}$ and $84\mathrm{th}$ percentiles.
    }
    \label{tab_apx1:Models}
\end{table*}

\bsp	
\label{lastpage}
\end{document}